\documentclass{Interspeech}
\usepackage{graphicx}
\usepackage{booktabs} 
\usepackage{longtable} 
\usepackage{lscape}
\usepackage{adjustbox}

\usepackage[belowskip=-12pt, aboveskip=1pt]{caption}
\interspeechcameraready

\title{Collecting, Curating, and Annotating Good Quality Speech deepfake dataset for Famous Figures: Process and Challenges}

\author{Hashim}{Ali}
\author{Surya}{Subramani}
\author{Raksha}{Varahamurthy}
\author[equalcontribution]{Nithin}{Adupa}
\author[equalcontribution]{Lekha}{Bollinani}
\author{Hafiz}{Malik}

\affiliation[nocounter]{Department of Electrical and Computer Engineering}{University of Michigan}{USA}
\email{alhashim@umich.edu, suryasss@umich.edu, rakshav@umich.edu, adupa@umich.edu, lekhab@umich.edu, hafiz@umich.edu}

\usepackage{comment}

\begin{document}

\maketitle

\begin{abstract}
Recent advances in speech synthesis have introduced unprecedented challenges in maintaining voice authenticity, particularly concerning public figures who are frequent targets of impersonation attacks. This paper presents a comprehensive methodology for collecting, curating, and generating synthetic speech data for political figures and a detailed analysis of challenges encountered. We introduce a systematic approach incorporating an automated pipeline for collecting high-quality bonafide speech samples, featuring transcription-based segmentation that significantly improves synthetic speech quality. We experimented with various synthesis approaches; from single-speaker to zero-shot synthesis, and documented the evolution of our methodology. The resulting dataset comprises bonafide and synthetic speech samples from ten public figures, demonstrating superior quality with a NISQA-TTS naturalness score of 3.69 and the highest human misclassification rate of 61.9\%.

    
\end{abstract}

\section{Introduction}
The last few years have seen an exceptional increase in the realism of synthesized speech \cite{tan2021survey, ren2020fastspeech, casanova2022yourtts, wang2301neural, li2023zse}. This high quality of synthesized speech and the ability to distribute it through social media platforms are giving rise to manipulated information in the digital ecosystem. According to a Global Risk Report by the World Economic Forum, misinformation and disinformation are the most serious threats predicted over the next two years \cite{wef_global}. The report states that approximately three billion people are expected to participate in electoral polls across multiple countries over the next two years, however, the widespread use of misinformation and disinformation and the tools to disseminate may undermine the legitimacy of newly elected governments. This can result in political unrest ranging from violent protests and hate crimes to civil confrontation and terrorism. 

These concerns have already manifested themselves in several high-profile incidents. In 2022, a synthetic video portrayed President Zelenskii allegedly asking for military surrender \cite{ebaker_russian_2022}. Subsequently, in 2024, a fake audio purported to be from President Biden was used in an attempt to influence voter participation in the primary elections in New Hampshire \cite{elliott_biden_2024}. The scope of this threat became multi-national when London Mayor Sadiq Khan was targeted through fabricated audio content regarding Armistice Day observations \cite{Marianna_sadiq_2024}. As speech synthesis technologies advance in capability and accessibility, influential public figures face increasing exposure to voice spoofing attacks that can systematically manipulate public opinion. The development of robust audio spoofing detection systems has therefore become crucial. However, such systems require comprehensive, high-quality datasets containing authentic and synthetic speech samples from public figures. Creating these datasets presents unique challenges, particularly when dealing with prominent individuals whose voices are frequently targeted for manipulation. This underscores the urgent need for systematic approaches to building and maintaining audio spoofing detection datasets that can effectively protect high-profile individuals from voice-based impersonation attacks.

In this paper, we present a comprehensive methodology for collecting and generating synthetic speech data for high-profile political figures while documenting the challenges encountered and the solutions developed throughout the process. Our methodology emphasizes three key aspects: (1) comprehensive coverage of authentic speech in diverse real-world contexts, including political speeches, media interviews, and public statements; (2) systematic curation of high-quality audio samples that capture the distinctive vocal characteristics and speech patterns of each individual; and (3) creation of corresponding synthetic speech using multiple text-to-speech (TTS) systems to represent realistic spoofing scenarios. The process involves a carefully designed pipeline to collect and process speech samples. First, we identify and collect high-quality source material from publicly available videos, ensuring diverse speaking contexts and acoustic conditions. This is followed by rigorous preprocessing steps, including speaker diarization to isolate the target speaker, automated transcription of speech segments, audio quality assessment, and segmentation into chunks while preserving the natural flow of speech. For each authentic speech segment, we generate the corresponding synthetic speech using various TTS approaches. The resulting dataset is available on request at our lab datasets website\footnote{https://datasets.issflab.net}.

The remainder of this paper is organized as follows. Section \ref{exist_data} provides a description of the existing relevant audio anti-spoofing datasets and their limitations. Section \ref{data_method} describes the design considerations for audio data collection, the data collection pipeline, and the challenges faced. Section \ref{synthetic_speech} describes the process for generating synthetic speech samples and the corresponding challenges. Finally, Section \ref{data_stat} provides the statistics of the different datasets and their quality comparisons.

\section{Existing Audio Anti-Spoofing Datasets} \label{exist_data}
The research community has developed various datasets to advance the field of Audio Spoof Detection. We can broadly classify these datasets into two categories, based on their speaker characteristics and intended applications. The first category, General Purpose Speaker Datasets, comprises audio data of anonymous speakers in controlled environments, focusing on developing generic audio spoof detection systems. The second category, Identity-Specific Datasets, addresses the challenges of protecting public figures from targeted voice spoofing attacks.

\subsection{General-Purpose Speaker Datasets}
General-purpose speaker datasets have played a vital role in the advancement of audio spoofing detection research by providing standardized benchmarks to evaluate detection systems. The ASVspoof Challenge series \cite{wang2020asvspoof, liu_asvspoof_2023, yamagishi2021asvspoof, wang24_asvspoof} has emerged as the primary benchmark in this domain, systematically evolving to address increasingly sophisticated spoofing attacks. The latest iteration, ASVspoof 5 \cite{wang24_asvspoof}, represents a significant advancement by incorporating crowd-sourced speech data and a diverse range of deepfake attacks. It includes real-world speech extracted from the Multilingual Librispeech (MLS) English partition \cite{pratap2020mls}, which consists of audiobook recordings. Based on the ASVspoof series, the ASVSpoof Laundered Database (ASVSpoofLD) \cite{ali2024audio} is developed by passing audio files from the ASVSpoof19 LA eval partition through a series of laundering attacks (additive noise, reverberation, recompression, resampling, etc.), introducing additional complexity for detection systems. The DeepFake Audio Detection Dataset (DFADD) \cite{du2024dfadd} and CODECFake dataset \cite{xie2024codecfake} represent contemporary datasets specifically designed to evaluate detection systems against recent neural TTS architectures and codec-based neural speech synthesis methods. Both datasets are derived from the VCTK corpus, which comprises high-quality speech recordings collected in a controlled laboratory environment. The Multi-Language Audio Anti-spoofing Dataset (MLAAD) \cite{muller2024mlaad} extends the scope to cross-lingual scenarios, addressing the increasingly global nature of audio spoofing threats.

These datasets share several key characteristics that define their utility in audio anti-spoofing research, such as controlled recording environments, standardized evaluation protocols, and balanced attack representations. However, their focus on controlled conditions, anonymous speakers, and the use of read speech limits their applicability in scenarios requiring protection of specific individuals under real-world conditions.


\begin{figure}[t]
  \centering
  \includegraphics[width=\linewidth]{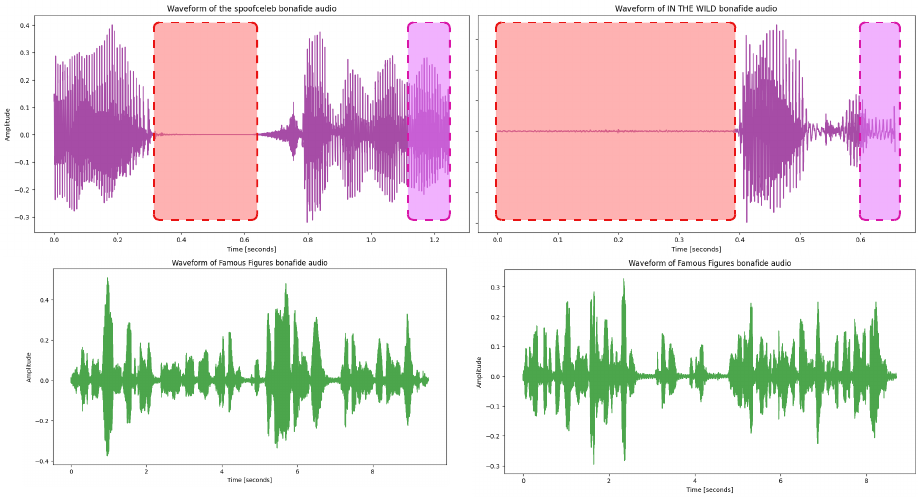}
  \caption{Speech clips from Spoofceleb \textbf{ (Top Left) } and In the Wild \textbf{ (Top Right)} with extended silence\textbf{ (Red tint) }, less duration and abrupt cut at the end \textbf{ (purple tint) }. The bottom two plots are from Famous figures dataset with an average duration of 8 seconds and Transcription based segments }
  \label{fig:speech_comparison}
\end{figure}
\subsection{Identity-Specific Datasets}
In contrast to general-purpose datasets, identity-specific datasets focus on protecting known individuals from targeted voice spoofing attacks. The In-The-Wild (ITW) dataset \cite{muller2022does} represents a unique collection of real-world speech data that bridges the gap between controlled laboratory evaluations and practical applications. Unlike the previous datasets, the ITW comprises 38 hours of speech data collected from various online platforms and social media sources. A significant advancement in identity-specific datasets is represented by SpoofCeleb \cite{jung2025spoofceleb}, which addresses several limitations of previous datasets by utilizing real-world data from VoxCeleb1 \cite{nagrani2020voxceleb}. The authors proposed a fully automated pipeline to process VoxCeleb1 speech samples and generate the corresponding synthetic speech.

Despite the use of genuine real-world audio samples, SpoofCeleb exhibits important limitations in the context of targeted speaker protection. Most notably, the dataset's training and evaluation partitions do not share common speakers, making it more suitable for generic deepfake detection rather than protecting specific individuals from targeted attacks. Moreover, Figure \ref{fig:speech_comparison} illustrates the data quality challenges in speech collection and their impact on the quality of the synthesis. The top waveforms are from SpoofCeleb (left) and In the Wild (right) datasets, which contain extended silences (red shading) and abrupt cuts (purple shading), which can lead to poor prosody and unnatural timing in synthetic speech. In contrast, the bottom waveforms show our Famous Figures dataset segments, which maintain an average duration of 8 seconds. 

\section{Dataset Design and Methodology} \label{data_method}
Our dataset design process is guided by three primary objectives: (1) ensuring high-quality authentic speech samples across various speaking contexts, (2) maintaining speaker diversity while capturing sufficient data per individual to represent their unique vocal characteristics, and (3) establishing a reproducible pipeline for data collection that can be extended to include additional public figures in the future. 

\subsection{Design Consideration} \label{design}
First, we established a criteria for selecting public figures based on three key factors: (a) frequency of public appearances, ensuring sufficient source material for data collection, (b) diversity of speaking contexts, including formal speeches, media interviews, and public statements, and (c) likelihood of being targeted for voice spoofing attacks based on their public influence. We selected 10 high-profile public figures who met these criteria. These figures include Anthony Blinken,  Barack Obama, Donald Trump, JD Vance, Joe Biden, Kamala Harris,  Mathew Miller, Tim Walz, Vivek Ramaswamy, and Elon Musk.  Second, we followed a systematic approach to the selection of the source material. We only collected YouTube videos which have (a) minimum resolution of 720p to ensure adequate audio quality, (b) minimum video duration of 5 minutes to ensure adequate speaking patterns, (c) publication date range of 2018-2024 to ensure current speaking styles, and (d) clear speech with minimal background noise.

\begin{figure}[t]
  \centering
  \includegraphics[width=\linewidth]{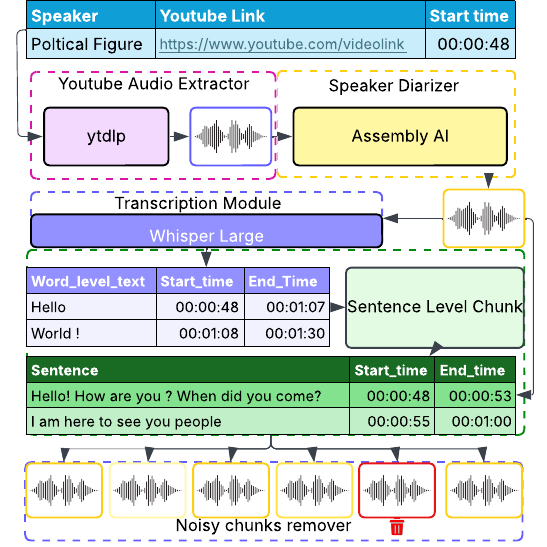}
  \caption{Schematic diagram for real audio data collection with transcription based segmentation.}
  \label{fig:data_collec_pipeline}
\end{figure}

\subsection{Data Collection Pipeline} \label{data_collection}
The data collection and processing pipeline is implemented as a systematic automated workflow to ensure consistency and reproducibility. The initial data acquisition begins with a carefully curated CSV file containing YouTube links to various speeches, interviews, and public appearances of selected public figures. Each entry in the CSV file includes metadata such as the speaker's identity, start time (the time at which the target speaker starts speaking), content type (speech, interview, etc.), publication date and the YouTube URL. As described in Figure \ref{fig:data_collec_pipeline}, the data collection pipeline consists of the following stages:

\begin{enumerate}
    \item \textbf{Audio Extraction:} We utilize yt-dlp\footnote{https://github.com/yt-dlp/yt-dlp} to directly download audio from YouTube links. It uses ffmpeg to download the best audio available in the WAV format, and resample it 16 kHz. Along with the YouTube video link, we specify the target speaker's speech starting time to trim the audio to begin from the specified timestamp, ensuring that the first speaker is the target speaker.

    \item \textbf{Speaker Diarization:} We employ Assembly AI\footnote{https://www.assemblyai.com/} to isolate segments that contain only the target speaker's voice. This step eliminates cross-talk and background speakers. 

    \item \textbf{Transcription Generation:} We integrate OpenAI Whisper Large Turbo\footnote{https://github.com/openai/whisper/} for transcription, which gave word-level transcripts with timestamps. We also experimented with the Google speech recognition api package and other commercial tools; however, they generated text with less accuracy and without proper punctuation.

    \item \textbf{Audio Segmentation:} We utilize transcription-based segmentation that predicts word-level transcripts along with their timestamps in the utterance, as illustrated in Figure \ref{fig:data_collec_pipeline}. This step processes word-level transcriptions with timestamps and groups them into sentence segments based on the utterance duration \( U \), user-defined duration \( D \), and the threshold duration \( T \). Words are sequentially appended to a segment, and when the duration of the segment reaches \( D - T \) seconds, the process searches for a punctuation mark within the interval \([D - T, D + 1]\). If a punctuation mark is found within this range, the segment is finalized up to the punctuation, while the remaining words are carried over to the next segment. If no punctuation is detected, the segment extends up to \( D + T \) seconds, with an additional 0.25 seconds of silence appended at the end. The last few words are discarded, if their cumulative duration is less than \( D - 2T \) seconds. In addition, incomplete sentences are discarded if the total segment count exceeds \( \left( \frac{U}{D} - 10 \right) \).

    \item \textbf{Quality Control:} Our initial segmentation strategy involved abrupt cutting at fixed intervals of \textit{n} seconds or minutes. For example, a 30-minute utterance would be divided into 300 segments of 6 seconds each. Through continuous experimentation, we transitioned to silence- or speaker-pause-based segmentation, as described in SpoofCeleb \cite{jung2025spoofceleb}. In this approach, if silence extends beyond 500 ms, the utterance is segmented at that point. We evaluated the output audio segments based on specific criteria, such as silence duration, sentence boundary completion, and SNR range. With the updated segmentation approach, we achieved segments with maximized voiced portions, completed sentence boundaries, and naturally reduced noise.
\end{enumerate}

\begin{table*}[t]
\centering
\caption{Statistical overview of audio deepfake datasets}
\label{tab:dataset_comparison}
\begin{tabular}{lccccll}
\toprule
Dataset & \#Speakers & \begin{tabular}[c]{@{}c@{}}Bonafide\\Utterances\end{tabular} & \begin{tabular}[c]{@{}c@{}}Synthetic\\Utterances\end{tabular} & \begin{tabular}[c]{@{}c@{}}Duration\\(hrs)\end{tabular} & Speaking Contexts & Synthesis Methods \\
\midrule
ASVspoof19 LA & 107 & 12,483 & 108,978 & $\sim$100 & Read speech & \begin{tabular}[c]{@{}l@{}}A01-A19 (TTS \& VC)\end{tabular} \\
ASVspoof 5 & 585 & 148,656 & 423,740 & $\sim$570 & \begin{tabular}[c]{@{}l@{}} Read speech\end{tabular} & \begin{tabular}[c]{@{}l@{}}44 TTS, VC systems\end{tabular} \\
DFADD & 109 & 44,455 & 163,500 & $\sim$50 & Read speech & \begin{tabular}[c]{@{}l@{}}Diffusion- and Flow-matching TTS\end{tabular} \\
CodecFake (EN) & 110 & 44,242 & 269,903 & $\sim$312 & Read speech & \begin{tabular}[c]{@{}l@{}} 7 Neural Audio Codecs\end{tabular} \\
SpoofCeleb & 1,251 & 248,000 & 2.5M+ & $\sim$400 & \begin{tabular}[c]{@{}l@{}} public appearances\end{tabular} & \begin{tabular}[c]{@{}l@{}}23 TTS systems \end{tabular} \\
In-The-Wild & 54 & 19,963 & 11,816 & $\sim$38 & \begin{tabular}[c]{@{}l@{}}public appearances\end{tabular} & Unknown \\
\midrule
Famous Figures & 10 & 26,500 & 265,000 & $\sim$590 & \begin{tabular}[c]{@{}l@{}}public appearances\end{tabular} & \begin{tabular}[c]{@{}l@{}}10 Open-source TTS models\end{tabular} \\
\bottomrule
\end{tabular}
\end{table*}

\begin{figure}[t]
  \centering
  \includegraphics[width=\linewidth]{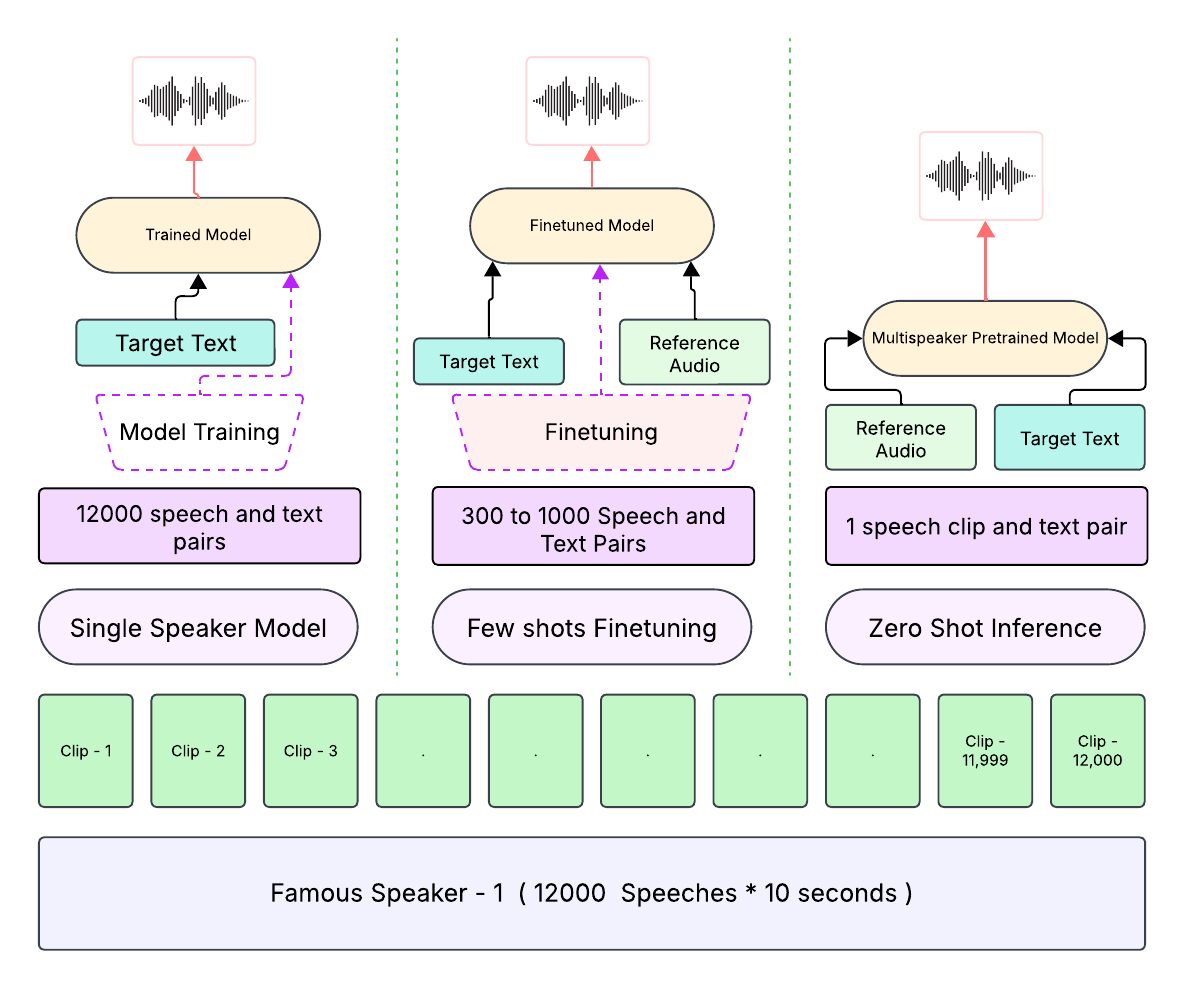}
  \caption{Schematic diagram of speech production.}
  \label{fig:speech_production}
\end{figure} 

\section{Synthetic Speech Generation} \label{synthetic_speech}

The evolution of synthetic speech has progressed from supervised neural models to self-supervised learning (SSL). Although early models produced robotic-sounding speech despite extensive training data, SSL-based methods significantly improved quality through large-scale pre-training. The recent integration of Audio Language Models (ALMs) has further enhanced prosody and expressiveness.


\subsection{Synthesis Pipeline} \label{process}
We first trained a total of 10 TTS models for each speaker using the audio samples and their corresponding transcriptions generated through the data collection pipeline in Section \ref{data_collection}. After that, we used the transcripts and the trained models to generate synthetic speech. From our exploration, we have identified three primary approaches to train TTS models.

\subsubsection{Speaker-Specific Training}
This approach involves training a model exclusively for a single speaker, which requires at least 24 hours of speech data (approximately 11,000 to 13,000 audio samples). The training process includes creating file lists using bonafide audio samples and their corresponding transcriptions, as mentioned in Figure \ref{fig:speech_production}. Training is performed using high-performance GPUs, such as three NVIDIA A100 GPUs, and typically takes five days per speaker. We train StyleTTS2 \cite{li2024styletts} only using this approach. The synthetic speech generated through this approach exhibited limitations in prosody and naturalness, which led us to explore more efficient alternatives.

\subsubsection{Few-Shot Fine-Tuning}
In this approach, a model pre-trained for multiple speakers is adapted to a specific speaker using 1 to 3 hours of its data through fine-tuning. We fine-tuned XTTSv2 \cite{casanova2024xtts} and StyleTTS2 \cite{li2024styletts} for all speakers. This approach significantly improved speech quality by effectively transferring prosody and emotional features from the multi-speaker model.


\subsubsection{Zero-Shot Synthesis}
In this approach, a large-scale model pre-trained for multiple speakers is adapted to a specific speaker using only a single reference audio and text pair. Although models like XTTSv2  and StyleTTS2 have zero-shot capabilities, they struggled to match the reference speaker’s voice accurately. However, recent models integrating ALM based architectures have drastically improved the performance. From this category, we have generated synthetic speech for all speakers using F5TTS\cite{chen2024f5}, E2TTS\cite{eskimez2024e2}, FishSpeech \cite{liao2024fish}, SSRSpeech \cite{wang2024ssr}, MaskGCT \cite{wang2024maskgct}, CozyVoice2 \cite{du2024cosyvoice2scalablestreaming}, LLASA,\cite{ye2025Vallescalingtraintimeinferencetime} and Zonosv0.1. 


\subsection{Speech Synthesis Challenges and Solutions} \label{synthesis_challenges}

The development of synthetic speech for political figures presented unique challenges that required iterative solutions. We discuss our progress through multiple approaches, highlighting both the challenges encountered and the solutions implemented.

\subsubsection{Challenges}
We faced significant obstacles in our initial attempts to use political speech recordings to train TTS models. We faced fundamental limitations with Signal-to-Noise Ratio (SNR) measurements for publicly available recordings (12.12dB), which is substantially below the 30.25 dB benchmark established by standard TTS datasets. The synthesis phase presented two primary challenges: maintaining speaker identity and ensuring natural-sounding output. Despite using various TTS (Tacotron-Capacitron\cite{wang2017tacotron}, GlowTTS\cite{kim2020glow}) and vocoder architectures (HiFi-GAN\cite{yang2023hifi}, UnivNet\cite{jang2021univnet}), we encountered issues with mechanical articulation and high-frequency noise.



\begin{table}[t]
    \centering
    \caption{NISQAv2 prediction for in the wild datasets}
    \begin{tabular}{lcc}
        \toprule
        Dataset type & Avg. Naturalness & Fake Miss-Rates(\%)\\
        \midrule
        ASVspoof19 LA & 2.99 & 25\\
        CodecFake (EN)&3.41 & 57.5\\
        DFADD & 3.39 & 24.4\\
        MLAAD & 3.53 & 34.8\\
        In the wild & 2.80 & 52.5\\
        Spoof celeb & 3.06 & - \\
        Famous Figures & \textbf{3.69} & \textbf{61.9}\\
        \bottomrule
    \end{tabular}
    \label{tab:pred_scores}
\end{table}

\subsubsection{Evolution of Solutions}

Our solution strategy evolved through three key phases:

\begin{enumerate}
    \item \textbf{Audiobook Data Approach:} We first attempted using high-quality audiobook data (5-10 hours per speaker) from Amazon Audible. While this improved signal quality and reduced computational artifacts, particularly with HiFi-GAN vocoder, the synthetic speech exhibited notable monotonicity, lacking the dynamic range essential for political discourse.
    
    \item \textbf{Enhanced Segmentation:} We implemented transcription-based sentence-level segmentation using Whisper Large Turbo's word-level timestamps as described in Section \ref{data_collection}. This approach preserved linguistic coherence and improved phoneme alignment, leading to reduced noise and more accurate spectrogram generation.
    
    \item \textbf{Advanced Model Architecture:} Finally, we transitioned to Few-Shot and Zero-Shot TTS models, which demonstrated superior performance compared to single-speaker training approaches, effectively addressing our remaining challenges.
\end{enumerate}

This iterative progression from traditional approaches to more sophisticated solutions ultimately enabled us to achieve higher quality synthetic speech while maintaining speaker-specific characteristics and natural prosody.


\section{Dataset Statistics and Analysis} \label{data_stat}

To evaluate the perceptual quality and detection difficulty of synthetic speech in datasets, we conducted subjective and objective assessments. For subjective evaluation, we implemented a web-based listening test with 32 unique participants. Each participant was presented with 14 randomly selected audio samples (two from each dataset, one real and one fake) and tasked with classifying them as either genuine or synthetic speech. The results revealed significant variations in detection difficulty across datasets. Notably, our Famous Figures dataset achieved the highest mis-classification rate at 61.9\%, suggesting that its synthetic speech samples more closely resemble natural speech compared to other datasets. The misclassification rates for different datasets are shown in Table \ref{tab:pred_scores} column Fake Miss-Rates.

For an objective quality assessment, we used the NISQA-TTS model \cite{mittag2021deep}, a deep learning-based model specifically designed to evaluate the quality of synthetic speech. As shown in Table \ref{tab:pred_scores}, our Famous Figures dataset achieved impressive quality scores, with NISQA-TTS predicting naturalness of 3.69, surpassing both In the Wild and Spoof Celeb datasets. 

\bibliographystyle{IEEEtran}
\bibliography{mybib}

\end{document}